\documentclass[twocolumn,preprintnumbers,amsmath,amssymb]{revtex4}

\usepackage{graphicx}
\usepackage{dcolumn}
\usepackage{bm}

\begin{document}


\title{Giant electrocaloric effect in thin film PbZr$_{0.95}$Ti$_{0.05}$O$_3$}

\author{A. Mischenko$^{1,}$\footnote{Electronic address: am507@cam.ac.uk}, Q. Zhang$^2$, J.F. Scott$^3$,
R.W. Whatmore$^2$, and N.D. Mathur$^1$}

\affiliation{$^{1)}$Department of Materials Science, Pembroke
Street, University of Cambridge, CB2 3QZ, Cambridge, United Kingdom}

\affiliation{$^{2)}$Department of Advanced Materials, SIMS,
Cranfield University, Beds MK43 0AL, United Kingdom}

\affiliation{$^{3)}$Centre for Ferroics, Earth Sciences Department,
University of Cambridge, Cambridge CB2 3EQ, United Kingdom}


\begin{abstract}
An applied electric field can reversibly change the temperature of
an electrocaloric material under adiabatic conditions, and the
effect is strongest near phase transitions. This phenomenon has been
largely ignored because only small effects~\cite{1TuttlePayne}
(0.003 K V$^{-1}$) have been seen in bulk samples such as
Pb$_{0.99}$Nb$_{0.02}$(Zr$_{0.75}$Sn$_{0.20}$Ti$_{0.05}$)$_{0.98}$O$_3$
and there is no consensus on macroscopic models~\cite{2ScottEncycl}.
Here we demonstrate a giant electrocaloric effect (0.48 K V$^{-1}$)
in 300 nm sol-gel PbZr$_{0.95}$Ti$_{0.05}$O$_3$
films~\cite{3Sawaguchi} near the ferroelectric Curie temperature of
222$^o$C. We also discuss a solid state device concept for
electrical refrigeration that has the capacity to outperform Peltier
or magnetocaloric coolers. Our results resolve the controversy
surrounding macroscopic models of the electrocaloric effect and may
inspire $ab~initio$ calculations of electrocaloric parameters and
thus a targeted search for new materials.
\end{abstract}

\maketitle

\section{\label{sec:Intro}Introduction}

There has been increasing interest in novel cooling technologies
over the last decades for several reasons. Firstly, it is important
to reduce greenhouse gases that are used heavily in domestic and
industrial refrigeration. Secondly, higher current densities in
integrated circuits will impose higher demands on cooling systems
that cannot be exclusively met by the current fan-based solutions.
Thirdly, new requirements such as fibre optic lasers are
continuously emerging. The electrocaloric (EC) effect in
ferroelectrics generated great interest~\cite{1TuttlePayne,
4Thacher, 5Lawless, 6Morrow} primarily in the 1950s-70s, but it has
not been exploited commercially.

Small EC effects were reported in previous works by direct and
indirect measurements that were in close
agreement~\cite{1TuttlePayne, 4Thacher, 7TuttleThesis}. For example,
bulk
Pb$_{0.99}$Nb$_{0.02}$(Zr$_{0.75}$Sn$_{0.20}$Ti$_{0.05}$)$_{0.98}$O$_3$
shows the highest EC effect measured so far, and the direct
measurements of the peak value (2.5 K in 30 kV cm$^{-1}$) were only
10-15\% less than the indirect measurements~\cite{1TuttlePayne,
7TuttleThesis}. This work reports our discovery of a giant EC effect
of 12~K in antiferroelectric thin films of Zr rich Pb(Zr,Ti)O$_3$
(PZT).

Microscopic theoretical treatments of the EC effect have not been
forthcoming because the macroscopic theory is unestablished. Three
leading textbooks on ferroelectricity differ on the macroscopic
physics of the EC effect~\cite{8Fatuzzo, 9Mitsui, 10Jona}. Fatuzzo
and Merz~\cite{8Fatuzzo} argue that the EC effect only occurs above
the phase transition (Curie) temperature $T_C$, where the
polarization $P$ is finite in the presence of the applied electric
field $E$. Mitsui, Tatsuzaki and Nakamura~\cite{9Mitsui} argue that
the EC effect can only occur below $T_C$, where the spontaneous
value of $P$ changes with temperature. Jona and
Shirane~\cite{10Jona} disagree with both~\cite{8Fatuzzo}
and~\cite{9Mitsui}, and assert that the effect occurs both above and
below $T_C$ but is larger above. More generally, microscopic models
of ferroelectrics are not well established. For example, it is only
in the last decade that a fully quantum mechanical treatment has
been available~\cite{11Resta, 12Kingsmith}.

We emphasise that there are no atomic-level $ab~initio$ models for
the electrocaloric effect. This may be partly because of the
confused macroscopic physics discussed above, and also because it
has been difficult to carry out $ab~initio$ calculations for finite
fields. However, in the past few months, Wu
$et~al.$~\cite{13Vanderbilt} have managed to treat some materials,
including BaTiO$_3$, in finite fields at absolute zero. It is our
hope that the present work will encourage $ab~initio$ calculations
of the electrocaloric effect. This will require further improvements
because both finite temperatures and fields are necessarily
involved.

\section{\label{sec:PZT}Zirconium rich PZT thin films}

It is plausible that antiferroelectric thin films of Zr-rich PZT
show promising EC effects, since the converse effect of
pyroelectricity is pronounced and forms the basis of infrared
detectors~\cite{14ZhangWhatmore}. Both Zr-rich PZT and the more
common compositions such as PbZr$_{0.48}$Ti$_{0.52}$O$_3$ are used
both as capacitors due to their high dielectric
constants~\cite{15Pan}, and also as high-strain
actuators/transducers and prototype microelectromechanical systems
due to their piezoelectric properties~\cite{15Pan}. However, the
potential for antiferroelectric films in cooling applications has
not been considered.

Bulk Pb(Zr$_{0.95}$Ti$_{0.05}$)O$_3$ is an orthorombic
antiferroelectric at room temperature. On heating to $\sim$120$^o$C,
this structure transforms to a rhombohedral ferroelectric phase.
There is substantial thermal hysteresis in this antiferroelectric to
ferroelectric transition, which on cooling occurs at $\sim$80$^o$C.
The structure transforms to cubic paraelectric above 242$^o$C. This
is a first-order phase transition with a Curie-Weiss temperature,
extrapolated linearly from the inverse dielectric susceptibility, of
$T_o$=225$^o$C~\cite{16WhatmoreThesis}. The rhombohedal to
paraelectric transition at this composition is close to a
tricritical point at PbZr$_{0.94}$Ti$_{0.06}$O$_3$ where its
character changes from first to second
order~\cite{17WhatmoreClarke}. These transition temperatures, which
were taken from single crystals and high purity ceramics, differ
somewhat from much earlier data~\cite{3Sawaguchi}, in which the
sample purity was not as great. The EC effect could not be predicted
from the literature, since there is no data for Zr-rich PZT thin
films at the high temperatures and high electric fields of interest.

\section{\label{sec:Experimental}Experimental}

PZT sols were prepared from Sigma-Aldrich precursors.
Pb(OAc)$_2$·3H$_2$O was dissolved in methanol and refluxed for
2~hours at~70$^o$C. Separately, a mixture of acetic acid and
methanol was added to a mixture of~Zr(O$^n$Pr)$_4$
and~Ti(O$^n$Bu)$_4$ and the resulting solution was stirred at room
temperature for two hours. The~Pb and~Zr/Ti based solutions were
mixed with a~20\% excess of the former to compensate for~Pb loss
during sintering. After gentle stirring, the yellow solution
obtained was passed through a~0.2~$\mu$m filter and stabilized by
the addition of ethylene glycol.

Sols were spin-coated at 3000~rpm for 30~s onto
Pt(111)/TiO$_x$/SiO$_2$/Si(100) substrates that had been rinsed with
acetone and propanol. Layers of $\sim$70~nm were obtained by
pre-firing in air on a hotplate at 300$^o$C for 60~s, and then
further annealing on another hotplate at 650$^o$C for 10~minutes.
This procedure was repeated five times to obtain $\sim$300~nm films.

Film structure was determined by x-ray diffraction on a Philips
diffractometer using Cu K$\alpha$  radiation. $\theta$-2$\theta$
scans corresponded to a polycrystalline perovskite phase with no
preferred orientations, and no traces of pyrochlore. Pt top
electrodes of diameter 0.2~$\mu$m were sputtered deposited through a
mechanical mask, and the bottom Pt electrode was contacted with
silver dag at a substrate edge. The dielectric constant and loss
tangent were measured using a HP 4192A Impedance Analyser at 100~kHz
and 100~mV ac amplitude. Hysteresis measurements were carried out at
10~kHz using a Radiant Technologies Precision Premier workstation
and a high temperature (280$^o$C) probe station. The temperature of
the sample was controlled via feedback from a thermocouple, accurate
to 0.3$^o$C, in contact with the sample.

\section{Results and interpretation}

Electrical hysteresis measurements were made roughly every
$T$~=~15$^o$C in the temperature range 35~-~280$^o$C, on cooling to
minimise reductions in $P$ due to fatigue. Representative plots
(Fig. 1) of $P(E)$ show the expected antiferroelectric
behaviour~\cite{3Sawaguchi} at 35$^o$C, and at 280$^o$C lossy
paraelectric behaviour that in these measurements is
indistinguishable from the ferroelectric loops at intermediate
temperatures. The dielectric constant $\varepsilon$ and loss tangent
each show a broad peak associated with the
ferroelectric-paraelectric transition at $T_C$=222$^o$C on cooling
(upper inset, Fig. 1) and $T_C$=215$^o$C on heating, but no peaks
corresponding to the antiferroelectric-ferroelectric transition can
be resolved. This broadness is typical of thin films and is likely
due to interfacial strain, scalar concentration gradients, or other
forms of microscopic variability~\cite{18Saad}.
\begin{figure}
\includegraphics[scale=0.8]{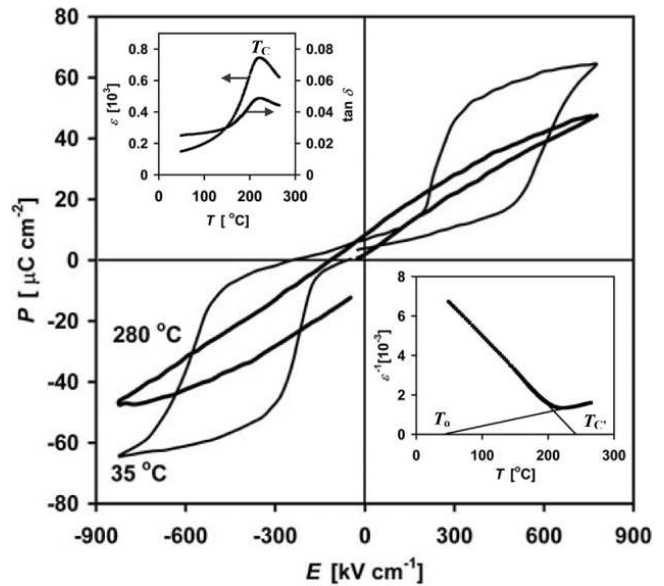}
\caption{\label{fig:Mischenko_Fig1} Electrical measurements of
Pb(Zr$_{0.95}$Ti$_{0.05}$)O$_3$ films on cooling. Polarisation $P$
versus applied electric field $E$ at~10~kHz showing evidence of
antiferroelectricity at~35$^o$C that persists up to~190$^o$C. Above
this temperature, we observe no qualitative change corresponding to
the ferroelectric to paraelectric transition that is apparent from
the insets. The data at~280$^o$C shows lossy paraelectric behaviour.
Upper inset: the real part of the effective dielectric constant
measured at~100~kHz shows a single peak at the bulk phase change
temperature $T_C$~=~222$^o$C, below which the
antiferroelectric-ferroelectric transition cannot be resolved. Loss
tangent tan~$\delta$~=~3\% at~35$^o$C. Lower inset: extrapolation of
the low temperature 1/$\varepsilon$  data to zero gives
$T_{C'}$~=~242$^o$C, and the corresponding high temperature
extrapolation gives the Curie-Weiss temperature
$T_o~\approx$~40$^o$C. Since $T_o$~$<$~$T_{C'}$, the
ferroelectric-paraelectric transition is first
order~\cite{19ScottMemories}.}
\end{figure}
\begin{figure}
\includegraphics[scale=1.7]{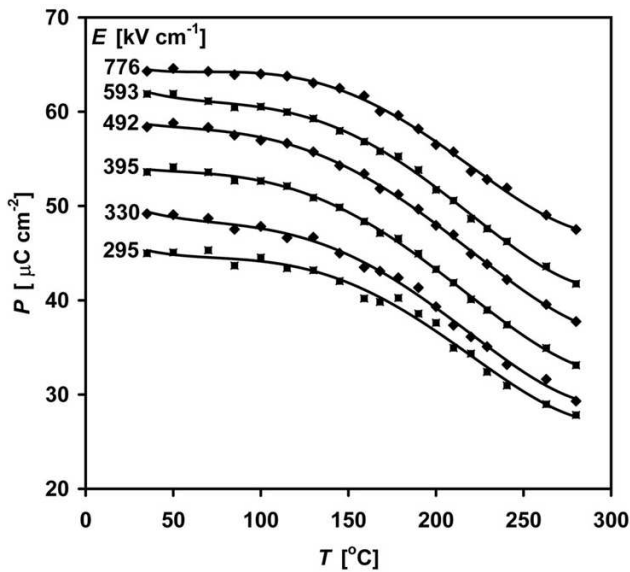}
\caption{\label{fig:Mischenko_Fig2} The temperature dependence of
polarisation $P$ at selected applied fields $E$. Data extracted for
$E$~$>$~0 from the upper branches of nine hysteresis loops measured
at 10~kHz in 35$^o$C~$\leq$~$T$~$\leq$~280$^o$C. The EC effect is
largest when $|\partial P/ \partial T|$ is maximised at the broad
paraelectric to ferroelectric transition. The lines represent
4$^{th}$ order polynomial fits to the data.}
\end{figure}
Reversible adiabatic changes in EC entropy $S$, energy $U$ and
temperature $T$ for a material of density $\rho$ with heat capacity
$C$ are given~\cite{7TuttleThesis} by:
\begin{eqnarray}
\Delta S &=& -\frac{1}{\rho}\int_{E_1}^{E_2}{\left(\frac{\partial
P}{\partial T}\right)_E}dE, \label{eq:DS} \\
\Delta U &=& -\frac{1}{\rho}\int_{E_1}^{E_2}{T\left(\frac{\partial
P}{\partial T}\right)_E}dE, \quad \textrm{and} \label{eq:DU}\\
\Delta T &=&
-\frac{1}{\rho}\int_{E_1}^{E_2}{\frac{T}{C}\left(\frac{\partial
P}{\partial T}\right)_E}dE,\label{eq:DT}
\end{eqnarray}
assuming the Maxwell relation  $\partial P / \partial T = \partial S
/ \partial E$. Values of $\partial P / \partial T$ were obtained
from 4$^{th}$ order polynomial fits to $P(T)$ data (Fig. 2). Fatigue
may only reduce our values of $|\partial P / \partial T|$ since the
data were taken on cooling such that~$P$ increased in successive
hysteresis measurements. In the temperature range of interest, the
heat capacity $C$~=~330~J~K$^{-1}$ kg$^{-1}$ remains sensibly
constant for Zr rich PZT films, and the peak associated with the
transition is $<$~10\% of the background~\cite{20DavitadzeThesis,
21Suchanek}. We note that assuming a constant value of $C$ despite a
$\approx$~50\% peak~\cite{7TuttleThesis} resulted in excellent
agreement with direct EC measurements of $T$ in bulk
Pb$_{0.99}$Nb$_{0.02}$(Zr$_{0.75}$Sn$_{0.20}$Ti$_{0.05}$)$_{0.98}$O$_3$~\cite{1TuttlePayne}.
A value of $\rho$~=~8.3~g~cm$^{-3}$ reported for the similar
compound (Pb,Zr,Sn)TiO$_3$ was used here~\cite{7TuttleThesis}. The
lower integration limit $E_1$~=~295~kV~cm$^{-1}$ was set
deliberately high to avoid the antiferroelectric regime (at low
fields, Fig. 1), which ensures that $\partial P / \partial T$~$<$~0.
The upper integration limit $E_2$~=~776~kV~cm$^{-1}$ represents the
maximum field at which a consistent dataset could be obtained.

EC entropy changes associated with the ferroelectric-paraelectric
transition are shown in Fig. 3. At  $E$~=~30~kV~cm$^{-1}$, the peak
value of $S$ is consistent with bulk values of
0.6~J~K$^{-1}$~kg$^{-1}$ in
Pb$_{0.99}$Nb$_{0.02}$(Zr$_{0.455}$Sn$_{0.455}$Ti$_{0.09}$)$_{0.98}$O$_3$~\cite{4Thacher}.
At larger fields, the peak value of $S$ is significantly larger, and
is comparable with values of $\approx$15~J~K$^{-1}$~kg$^{-1}$
reported for the best magnetocaloric materials in large magnetic
fields of 5~T~\cite{22Pecharsky, 23Tegus, 24Bruck}. Since hysteresis
losses are at worst 5\% of the EC energy changes (Fig. 3, inset),
the reversible thermodynamics assumed here (equations~1-3) is a
reasonable approximation.

\begin{figure}
\includegraphics[scale=0.8]{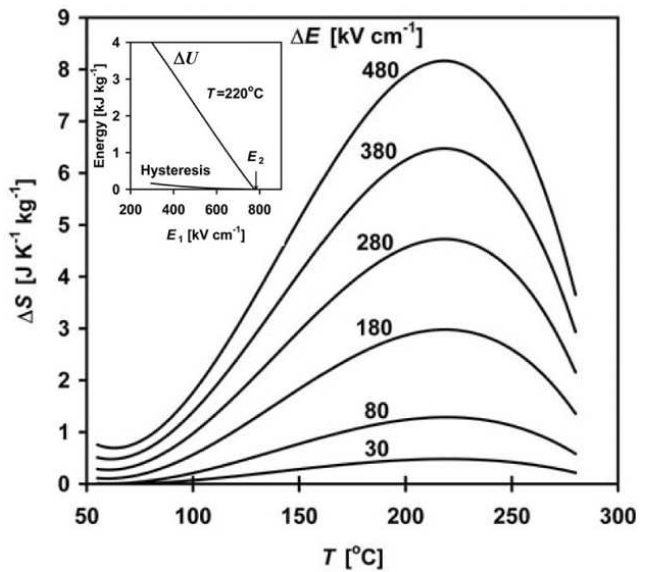}
\caption{\label{fig:Mischenko_Fig3} Electrocaloric entropy and
energy changes. Main panel, entropy changes $S$ were calculated from
equation~(1) using the fits presented in Fig.
\ref{fig:Mischenko_Fig1}, at selected values of  $E~=~E_2~-~E_1$
with $E_2$~=~776~kV~cm$^{-1}$. Inset, at $T$~=~220$^o$C the EC
energy~$U$ was similarly calculated from equation~(2) as a function
of $E~=~E_2~-~E_1$, with $E_2$~=~776~kV~cm$^{-1}$. The corresponding
hysteresis losses as determined from the $P(E)$ loops (e.g.
Fig.~\ref{fig:Mischenko_Fig1}) are small. This justifies the
indirect approach used to calculate the strength of the EC effect.}
\end{figure}

\begin{figure}
\includegraphics[scale=1.7]{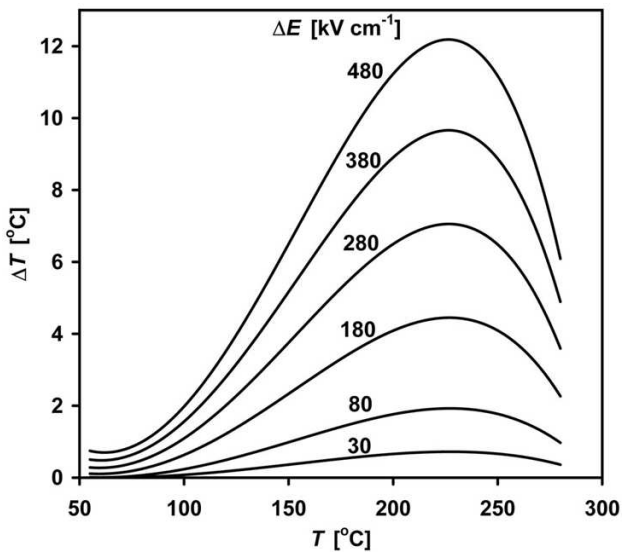}
\caption{\label{fig:Mischenko_Fig4} Electrocaloric temperature
changes.  $T$ was calculated from equation~(3) using selected values
of $E~=~E_2~-~E_1$, with $E_2$~=~776~kV~cm$^{-1}$. The peak value
of~12$^o$C occurs at $T_{EC}$~=~226$^o$C with
$E$~=~480~kV~cm$^{-1}$.}
\end{figure}

EC temperature changes corresponding to the entropy changes in Fig.
3 are presented in Fig. 4. The largest change (12$^o$C in 25~V, i.e.
0.48~K~V$^{-1}$) at $T_{EC}$~=~226$^o$C dwarfs the previous best
results obtained in bulk
Pb$_{0.99}$Nb$_{0.02}$(Zr$_{0.75}$Sn$_{0.20}$Ti$_{0.05}$)$_{0.98}$O$_3$
(2.5$^o$C in 750~V, i.e. 0.003~K~V$^{-1}$) at $T_{EC}$~=~162$^o$C.
At all electric fields studied, the effects persist both above and
below $T_C$=222$^o$C, unlike the models of~\cite{8Fatuzzo}
and~\cite{9Mitsui}, possibly because the transitions are not sharp
in our films. The effect peaks at $T_C<T_{EC}$ supporting the model
of~\cite{10Jona}, and since $T_{C'}$~=~242$^o$C (lower inset, Fig.
1), we also have $T_C<T_{EC}<T_{C'}$ which implies that our data
support the picture for first order phase transitions presented
in~\cite{2ScottEncycl}.

We propose that the EC effect demonstrated here in
Pb(Zr$_{0.95}$Ti$_{0.05}$)O$_3$ films may be exploited in an
efficient solid state heat pump that two of us (A.M. and N.D.M.)
proposed previously~\cite{25MischenkoMathur1, 26MischenkoMathur2}.
In this design, thermoelectric heat switches such as Peltier units
separate the EC element from the heat source and heat sink. Each
switch is open when driven by a sufficiently large forward current.
In the absence of this current the switch is closed and conducts
heat passively. A reverse current enhances heat flow. Multiple EC
elements with a range of working temperatures may be separated by
heat switches in order to span larger temperature ranges.

A future challenge is to increase cooling power by increasing EC
film thickness. Simulations assuming bulk
Pb$_{0.99}$Nb$_{0.02}$(Zr$_{0.75}$Sn$_{0.20}$Ti$_{0.05}$)$_{0.98}$O$_3$
properties~\cite{7TuttleThesis} represent lower bounds on
performance. For a 6~K change across four such
30$\times$30$\times$2~mm$^3$ EC elements, separated by
30$\times$30$\times$0.5~mm$^3$ Bi$_2$Te$_3$ Peltier heat switches,
we obtained at a device operating frequency of 10~Hz a coefficient
of performance approaching $\approx$50\% of the Carnot limit, i.e. a
cooling power of $\approx$50~W for an input of several watts.

We now make a brief comparison with competing refrigeration
technologies. Bulk magnetocaloric materials such as
Gd$_5$Ge$_2$Si$_2$ (15~K in 5~T)~\cite{22Pecharsky} and
MnFeP$_{0.45}$As$_{0.55}$ (4.2~K in 1.45~T)~\cite{24Bruck} show
impressive effects, and the efficiency of a prototype cooling system
is~60\% of the Carnot limit~\cite{28Zimm}. However, the large
magnetic fields required are impractical, whereas by contrast it is
relatively simple to generate the large electric fields relevant to
the EC effect. Electrically driven Peltier coolers~\cite{29Goldsmid}
based on Bi$_2$Te$_3$ semiconductors are compact and cheap, but
show~\cite{30Otey} poor coefficients of performance, e.g. 3\% for a
cooling power of 15~W with  $\Delta T$~=~10~K.

The EC effect demonstrated here will hopefully inspire theoretical
insights into ferroelectrics, materials improvements that involve
e.g. doping~\cite{14ZhangWhatmore, 31Zhang} or switching to
bismuth-based ferroelectrics~\cite{32Fu}, and ultimately commercial
applications. For example, the EC effect could provide cooling
solutions for electronic components such as computer chips. The
pyroelectric effect could be used for example to recover useful
electrical power from waste heat. Indeed, we note that the
optimisation of one effect necessarily optimises the other, and that
stress free environments should be more attractive than the strain
free environments experienced by the thin films.\\

\begin{acknowledgments}
A.M. was supported by Churchill College, Cambridge, a Kapitza
Scholarship from the Cambridge Overseas Trust, and an ORS award from
Universities UK. N.D.M. thanks The Royal Society for support.
Cranfield University gratefully acknowledge financial support from
EPSRC under the Platform Grant~GR/R92448/01. We thank F.D.~Morrison
and D.-J.~Jung for discussions.
\end{acknowledgments}

\bibliography{MischenkoECE}

\begin{thebibliography}{31}
\expandafter\ifx\csname natexlab\endcsname\relax\def\natexlab#1{#1}\fi
\expandafter\ifx\csname bibnamefont\endcsname\relax
  \def\bibnamefont#1{#1}\fi
\expandafter\ifx\csname bibfnamefont\endcsname\relax
  \def\bibfnamefont#1{#1}\fi
\expandafter\ifx\csname citenamefont\endcsname\relax
  \def\citenamefont#1{#1}\fi
\expandafter\ifx\csname url\endcsname\relax
  \def\url#1{\texttt{#1}}\fi
\expandafter\ifx\csname urlprefix\endcsname\relax\def\urlprefix{URL }\fi
\providecommand{\bibinfo}[2]{#2}
\providecommand{\eprint}[2][]{\url{#2}}

\bibitem[{\citenamefont{Tuttle and Payne}(1981)}]{1TuttlePayne}
\bibinfo{author}{\bibfnamefont{B.~A.} \bibnamefont{Tuttle}} \bibnamefont{and}
  \bibinfo{author}{\bibfnamefont{D.}~\bibnamefont{Payne}},
  \bibinfo{journal}{Ferroelectrics} \textbf{\bibinfo{volume}{37}},
  \bibinfo{pages}{603} (\bibinfo{year}{1981}).

\bibitem[{\citenamefont{Scott}(1993)}]{2ScottEncycl}
\bibinfo{author}{\bibfnamefont{J.}~\bibnamefont{Scott}},
  \bibinfo{journal}{Encyclopedia of applied physics, VCH, Berlin}
  \textbf{\bibinfo{volume}{5}}, \bibinfo{pages}{1} (\bibinfo{year}{1993}).

\bibitem[{\citenamefont{Sawaguchi}(1953)}]{3Sawaguchi}
\bibinfo{author}{\bibfnamefont{E.}~\bibnamefont{Sawaguchi}},
  \bibinfo{journal}{J. Phys. Soc. Japan} \textbf{\bibinfo{volume}{8}},
  \bibinfo{pages}{615} (\bibinfo{year}{1953}).

\bibitem[{\citenamefont{Thacher}(1968)}]{4Thacher}
\bibinfo{author}{\bibfnamefont{P.~D.} \bibnamefont{Thacher}},
  \bibinfo{journal}{J. Appl. Phys.} \textbf{\bibinfo{volume}{39}},
  \bibinfo{pages}{1996} (\bibinfo{year}{1968}).

\bibitem[{\citenamefont{Lawless}(1977)}]{5Lawless}
\bibinfo{author}{\bibfnamefont{W.~N.} \bibnamefont{Lawless}},
  \bibinfo{journal}{Phys. Rev. B.} \textbf{\bibinfo{volume}{16}},
  \bibinfo{pages}{433} (\bibinfo{year}{1977}).

\bibitem[{\citenamefont{Morrow and Lawless}(1977)}]{6Morrow}
\bibinfo{author}{\bibfnamefont{A.~J.} \bibnamefont{Morrow}} \bibnamefont{and}
  \bibinfo{author}{\bibfnamefont{W.~N.} \bibnamefont{Lawless}},
  \bibinfo{journal}{Ferroelectrics} \textbf{\bibinfo{volume}{15}},
  \bibinfo{pages}{159} (\bibinfo{year}{1977}).

\bibitem[{\citenamefont{Tuttle}(1981)}]{7TuttleThesis}
\bibinfo{author}{\bibfnamefont{B.~A.} \bibnamefont{Tuttle}}, Ph.D. thesis,
  \bibinfo{school}{University of Illinois at Urbana - Champaign}
  (\bibinfo{year}{1981}).

\bibitem[{\citenamefont{Fatuzzo and Merz}(1967)}]{8Fatuzzo}
\bibinfo{author}{\bibfnamefont{E.}~\bibnamefont{Fatuzzo}} \bibnamefont{and}
  \bibinfo{author}{\bibfnamefont{W.~J.} \bibnamefont{Merz}},
  \emph{\bibinfo{title}{Ferroelectricity}} (\bibinfo{publisher}{North-Holland,
  Amsterdam}, \bibinfo{year}{1967}).

\bibitem[{\citenamefont{Mitsui}(1976)}]{9Mitsui}
\bibinfo{author}{\bibfnamefont{T.}~\bibnamefont{Mitsui}},
  \emph{\bibinfo{title}{Introduction to the Physics of Ferroelectricity}}
  (\bibinfo{publisher}{Gordon and Breach, London}, \bibinfo{year}{1976}).

\bibitem[{\citenamefont{Jona and Shirane}(1962)}]{10Jona}
\bibinfo{author}{\bibfnamefont{F.}~\bibnamefont{Jona}} \bibnamefont{and}
  \bibinfo{author}{\bibfnamefont{G.}~\bibnamefont{Shirane}},
  \emph{\bibinfo{title}{Ferroelectric Crystals}} (\bibinfo{publisher}{McMillan,
  NY}, \bibinfo{year}{1962}).

\bibitem[{\citenamefont{Resta}(1994)}]{11Resta}
\bibinfo{author}{\bibfnamefont{R.}~\bibnamefont{Resta}}, \bibinfo{journal}{Rev.
  Mod. Phys.} \textbf{\bibinfo{volume}{66}}, \bibinfo{pages}{899}
  (\bibinfo{year}{1994}).

\bibitem[{\citenamefont{Kingsmith and Vanderbilt}(1993)}]{12Kingsmith}
\bibinfo{author}{\bibfnamefont{R.~D.} \bibnamefont{Kingsmith}}
  \bibnamefont{and}
  \bibinfo{author}{\bibfnamefont{D.}~\bibnamefont{Vanderbilt}},
  \bibinfo{journal}{Phys. Rev. B.} \textbf{\bibinfo{volume}{47}},
  \bibinfo{pages}{1651} (\bibinfo{year}{1993}).

\bibitem[{\citenamefont{Wu et~al.}(2005)\citenamefont{Wu, Vanderbilt, and
  Hamann}}]{13Vanderbilt}
\bibinfo{author}{\bibfnamefont{X.}~\bibnamefont{Wu}},
  \bibinfo{author}{\bibfnamefont{D.}~\bibnamefont{Vanderbilt}},
  \bibnamefont{and} \bibinfo{author}{\bibfnamefont{D.~R.}
  \bibnamefont{Hamann}}, \bibinfo{journal}{Phys. Rev. B.}
  \textbf{\bibinfo{volume}{72}}, \bibinfo{pages}{035105}
  (\bibinfo{year}{2005}).

\bibitem[{\citenamefont{Zhang and Whatmore}(2003)}]{14ZhangWhatmore}
\bibinfo{author}{\bibfnamefont{Q.}~\bibnamefont{Zhang}} \bibnamefont{and}
  \bibinfo{author}{\bibfnamefont{R.~W.} \bibnamefont{Whatmore}},
  \bibinfo{journal}{J. Appl. Phys.} \textbf{\bibinfo{volume}{94}},
  \bibinfo{pages}{5228} (\bibinfo{year}{2003}).

\bibitem[{\citenamefont{Pan et~al.}(1989)\citenamefont{Pan, Dam, Zhang, and
  Cross}}]{15Pan}
\bibinfo{author}{\bibfnamefont{W.~Y.} \bibnamefont{Pan}},
  \bibinfo{author}{\bibfnamefont{C.~Q.} \bibnamefont{Dam}},
  \bibinfo{author}{\bibfnamefont{Q.~M.} \bibnamefont{Zhang}}, \bibnamefont{and}
  \bibinfo{author}{\bibfnamefont{L.~E.} \bibnamefont{Cross}},
  \bibinfo{journal}{J. Appl. Phys.} \textbf{\bibinfo{volume}{66}},
  \bibinfo{pages}{6014} (\bibinfo{year}{1989}).

\bibitem[{\citenamefont{Whatmore}(1977)}]{16WhatmoreThesis}
\bibinfo{author}{\bibfnamefont{R.}~\bibnamefont{Whatmore}}, Ph.D. thesis,
  \bibinfo{school}{University of Cambridge} (\bibinfo{year}{1977}).

\bibitem[{\citenamefont{Whatmore et~al.}(1978)\citenamefont{Whatmore, Clarke,
  and Glazer}}]{17WhatmoreClarke}
\bibinfo{author}{\bibfnamefont{R.}~\bibnamefont{Whatmore}},
  \bibinfo{author}{\bibfnamefont{R.}~\bibnamefont{Clarke}}, \bibnamefont{and}
  \bibinfo{author}{\bibfnamefont{A.}~\bibnamefont{Glazer}},
  \bibinfo{journal}{J. Phys. C: Solid State Phys.}
  \textbf{\bibinfo{volume}{11}}, \bibinfo{pages}{3089} (\bibinfo{year}{1978}).

\bibitem[{\citenamefont{Saad et~al.}(2004)\citenamefont{Saad, Baxter, Bowman,
  Gregg, Morrison, and Scott}}]{18Saad}
\bibinfo{author}{\bibfnamefont{M.~M.} \bibnamefont{Saad}},
  \bibinfo{author}{\bibfnamefont{P.}~\bibnamefont{Baxter}},
  \bibinfo{author}{\bibfnamefont{R.~M.} \bibnamefont{Bowman}},
  \bibinfo{author}{\bibfnamefont{J.}~\bibnamefont{Gregg}},
  \bibinfo{author}{\bibfnamefont{F.~D.} \bibnamefont{Morrison}},
  \bibnamefont{and} \bibinfo{author}{\bibfnamefont{J.~F.} \bibnamefont{Scott}},
  \bibinfo{journal}{J. Phys.: Cond. Matt.} \textbf{\bibinfo{volume}{16}},
  \bibinfo{pages}{L451} (\bibinfo{year}{2004}).

\bibitem[{\citenamefont{Scott}(2000)}]{19ScottMemories}
\bibinfo{author}{\bibfnamefont{J.}~\bibnamefont{Scott}},
  \emph{\bibinfo{title}{Ferroelectric Memories}} (\bibinfo{publisher}{Springer,
  Berlin}, \bibinfo{year}{2000}).

\bibitem[{\citenamefont{Davitadze}(2003)}]{20DavitadzeThesis}
\bibinfo{author}{\bibfnamefont{S.}~\bibnamefont{Davitadze}}, Ph.D. thesis,
  \bibinfo{school}{Lomonosov Moscow State University} (\bibinfo{year}{2003}).

\bibitem[{\citenamefont{Suchanek et~al.}(2002)\citenamefont{Suchanek, Gerlach,
  Deyneka, Jastrabik, Davitadze, and Strukov}}]{21Suchanek}
\bibinfo{author}{\bibfnamefont{G.}~\bibnamefont{Suchanek}},
  \bibinfo{author}{\bibfnamefont{G.}~\bibnamefont{Gerlach}},
  \bibinfo{author}{\bibfnamefont{A.}~\bibnamefont{Deyneka}},
  \bibinfo{author}{\bibfnamefont{L.}~\bibnamefont{Jastrabik}},
  \bibinfo{author}{\bibfnamefont{S.}~\bibnamefont{Davitadze}},
  \bibnamefont{and} \bibinfo{author}{\bibfnamefont{B.}~\bibnamefont{Strukov}},
  \bibinfo{journal}{Mat. Res. Soc. Proc.} \textbf{\bibinfo{volume}{D8.4.1}},
  \bibinfo{pages}{718} (\bibinfo{year}{2002}).

\bibitem[{\citenamefont{Pecharsky and Gschneidner}(1997)}]{22Pecharsky}
\bibinfo{author}{\bibfnamefont{V.~K.} \bibnamefont{Pecharsky}}
  \bibnamefont{and} \bibinfo{author}{\bibfnamefont{K.~A.}
  \bibnamefont{Gschneidner}}, \bibinfo{journal}{Phys. Rev. Lett.}
  \textbf{\bibinfo{volume}{78}}, \bibinfo{pages}{4494} (\bibinfo{year}{1997}).

\bibitem[{\citenamefont{Tegus et~al.}(2002)\citenamefont{Tegus, Bruck, Buschow,
  and de~Boer}}]{23Tegus}
\bibinfo{author}{\bibfnamefont{O.}~\bibnamefont{Tegus}},
  \bibinfo{author}{\bibfnamefont{E.}~\bibnamefont{Bruck}},
  \bibinfo{author}{\bibfnamefont{K.~H.~J.} \bibnamefont{Buschow}},
  \bibnamefont{and} \bibinfo{author}{\bibfnamefont{F.~R.}
  \bibnamefont{de~Boer}}, \bibinfo{journal}{Nature}
  \textbf{\bibinfo{volume}{415}}, \bibinfo{pages}{150} (\bibinfo{year}{2002}).

\bibitem[{\citenamefont{Brück et~al.}(2005)\citenamefont{Brück, Ilyn, Tishin,
  and Tegus}}]{24Bruck}
\bibinfo{author}{\bibfnamefont{E.}~\bibnamefont{Brück}},
  \bibinfo{author}{\bibfnamefont{M.}~\bibnamefont{Ilyn}},
  \bibinfo{author}{\bibfnamefont{A.~M.} \bibnamefont{Tishin}},
  \bibnamefont{and} \bibinfo{author}{\bibfnamefont{O.}~\bibnamefont{Tegus}},
  \bibinfo{journal}{J. Magn. Magn. Mater.} \textbf{\bibinfo{volume}{290-291}},
  \bibinfo{pages}{8} (\bibinfo{year}{2005}).

\bibitem[{\citenamefont{Mischenko and Mathur}(2005)}]{25MischenkoMathur1}
\bibinfo{author}{\bibfnamefont{A.~S.} \bibnamefont{Mischenko}}
  \bibnamefont{and} \bibinfo{author}{\bibfnamefont{N.~D.}
  \bibnamefont{Mathur}}, in \emph{\bibinfo{booktitle}{APS March Meeting}}
  (\bibinfo{address}{Los Angeles, USA}, \bibinfo{year}{2005}).

\bibitem[{\citenamefont{Mischenko and Mathur}(2004)}]{26MischenkoMathur2}
\bibinfo{author}{\bibfnamefont{A.~S.} \bibnamefont{Mischenko}}
  \bibnamefont{and} \bibinfo{author}{\bibfnamefont{N.~D.}
  \bibnamefont{Mathur}}, \bibinfo{journal}{UK (0426230.9) and US (60/682,295)
  patents pending}  (\bibinfo{year}{2004}).

\bibitem[{\citenamefont{Zimm et~al.}(1998)\citenamefont{Zimm, Jastrab,
  Sternberg, Pecharsky, Gschneidner, Jr., and Anderson}}]{28Zimm}
\bibinfo{author}{\bibfnamefont{C.}~\bibnamefont{Zimm}},
  \bibinfo{author}{\bibfnamefont{A.}~\bibnamefont{Jastrab}},
  \bibinfo{author}{\bibfnamefont{A.}~\bibnamefont{Sternberg}},
  \bibinfo{author}{\bibfnamefont{V.}~\bibnamefont{Pecharsky}},
  \bibinfo{author}{\bibfnamefont{K.}~\bibnamefont{Gschneidner}},
  \bibinfo{author}{\bibfnamefont{M.}~\bibnamefont{Jr.},
  \bibfnamefont{Osborne}}, \bibnamefont{and}
  \bibinfo{author}{\bibfnamefont{I.}~\bibnamefont{Anderson}},
  \bibinfo{journal}{Adv. Cryog. Eng.} \textbf{\bibinfo{volume}{43}},
  \bibinfo{pages}{1759} (\bibinfo{year}{1998}).

\bibitem[{\citenamefont{Goldsmid}(1986)}]{29Goldsmid}
\bibinfo{author}{\bibfnamefont{H.~J.} \bibnamefont{Goldsmid}},
  \emph{\bibinfo{title}{Electronic refrigeration}} (\bibinfo{publisher}{Pion
  Ltd., London}, \bibinfo{year}{1986}).

\bibitem[{\citenamefont{Otey and Moskowitz}(2001)}]{30Otey}
\bibinfo{author}{\bibfnamefont{R.}~\bibnamefont{Otey}} \bibnamefont{and}
  \bibinfo{author}{\bibfnamefont{B.}~\bibnamefont{Moskowitz}},
  \bibinfo{journal}{SPIE magazine of photonics technologies and aplications}
  \textbf{\bibinfo{volume}{1, 3}}, \bibinfo{pages}{34} (\bibinfo{year}{2001}).

\bibitem[{\citenamefont{Zhang}(2004)}]{31Zhang}
\bibinfo{author}{\bibfnamefont{Q.}~\bibnamefont{Zhang}}, \bibinfo{journal}{J.
  Phys. D.: Appl. Phys.} \textbf{\bibinfo{volume}{37}}, \bibinfo{pages}{98}
  (\bibinfo{year}{2004}).

\bibitem[{\citenamefont{Fu et~al.}(2002)\citenamefont{Fu, Suzuki, and
  K.}}]{32Fu}
\bibinfo{author}{\bibfnamefont{D.}~\bibnamefont{Fu}},
  \bibinfo{author}{\bibfnamefont{K.}~\bibnamefont{Suzuki}}, \bibnamefont{and}
  \bibinfo{author}{\bibfnamefont{K.}~\bibnamefont{K.}}, \bibinfo{journal}{Jpn.
  J. App. Phys.} \textbf{\bibinfo{volume}{41}}, \bibinfo{pages}{L1103}
  (\bibinfo{year}{2002}).

\end{thebibliography}

\end{document}